\documentclass[runningheads,a4paper]{llncs}
\usepackage{amssymb}
\usepackage{graphicx}
\usepackage{listings}
\usepackage[tight]{subfigure}
\usepackage{wrapfig,booktabs}

\usepackage{xcolor}
\usepackage{listings}
\usepackage[hyphens]{url}
\usepackage[hidelinks]{hyperref}
\usepackage{float}

\usepackage{lipsum}
\usepackage{courier}
\usepackage{datetime}
\usepackage{graphicx}
\usepackage{stfloats}

\usepackage[T2A,T1]{fontenc}
\usepackage[utf8]{inputenc}
\usepackage[russian,english]{babel}
\usepackage{array}

\usepackage{newunicodechar}

\newunicodechar{ŝ}{\^{s}} 

\usepackage{lmodern}
\usepackage{algpseudocode} 

\usepackage{cite}
\urldef{\mailsa}\path|{mkelly, jbrunelle, mweigle, mln}@cs.odu.edu|    
\newcommand{\keywords}[1]{\par\addvspace\baselineskip
\noindent\keywordname\enspace\ignorespaces#1}
\usepackage{colortbl}
\usepackage{alltt}

\newcommand{\specialcell}[2][c]{%
  \begin{tabular}[#1]{@{}c@{}}#2\end{tabular}}

\begin{document}

\setlength{\textfloatsep}{0pt}
\setlength{\floatsep}{0pt}
\setlength{\intextsep}{4pt}
\setlength{\abovecaptionskip}{0pt}
\setlength{\belowcaptionskip}{4pt}
\setlength{\subfigcapskip}{0pt}

\mainmatter

\title{On the Change in Archivability of Websites Over Time}

\author{Mat Kelly, Justin F. Brunelle, Michele C. Weigle, and Michael L. Nelson}
\institute{Old Dominion University, Department of Computer Science\\
Norfolk VA, 23529, USA\\
\mailsa\\}

\maketitle

\begin{abstract}
As web technologies evolve, web archivists work to keep up so that our digital history is preserved. Recent advances in web technologies have introduced client-side executed scripts that load data without a referential identifier or that require user interaction (e.g., content loading when the page has scrolled). These advances have made automating methods for capturing web pages more difficult. Because of the evolving schemes of publishing web pages along with the progressive capability of web preservation tools, the \textit{archivability} of pages on the web has varied over time. In this paper we show that the archivability of a web page can be deduced from the type of page being archived, which aligns with that page's accessibility in respect to dynamic content. We show concrete examples of when these technologies were introduced by referencing mementos of pages that have persisted through a long evolution of available technologies. Identifying these reasons for the inability of these web pages to be archived in the past in respect to accessibility serves as a guide for ensuring that content that has longevity is published using good practice methods that make it available for preservation.
\keywords{Web Archiving, Digital Preservation}
\end{abstract}

\section{Introduction}
The web has gone through a gradient yet demarcated series of phases in which interactivity has become more fluid to the end-user. Early websites were static. Adoption of JavaScript allowed the components on a web page to respond to users' actions or be manipulated in ways that made the page more usable. Ajax \cite{garrett2005ajax} combines multiple web technologies to give web pages the ability to perform operations asynchronously. The adoption of Ajax by web developers facilitated the fluidity of user interaction on the web. Through each phase in the progression of the web, the ability to preserve the content displayed to the user has also progressed but in a less linear trend.

A large amount of the difficulty in web archiving stems from the crawler's insufficient ability to capture content related to JavaScript. Because JavaScript is executed on the client side (i.e., within the browser after the page has loaded), it should follow that the archivability could be evaluated using a consistent replay medium. The medium used to archive (normally a web crawler tailored for archiving, e.g., Heritrix \cite{mohr2004heritrix}) 
is frequently different from the medium used to replay the archive (henceforth, the \textit{web browser}, the predominant means of replay). The crawler creates the web archive, which is processed by a replay system (e.g., Internet Archive's Wayback Machine \cite{tofel2007wayback}), which is then accessed by an end-user through the web browser medium (e.g., the user accesses Wayback's web interface). This inconsistency between the perspective used to capture the pages versus the perspective used to view the stored content \cite{mythesis} introduces difficulty in evaluating the web resources' potential to be archived (henceforth the \textit{archivability}). Further discrepancies in the capabilities of the crawler versus the web browser make archivability difficult to measure without manual inspection. 

The success of preservation of a web page is defined by how much of the originally displayed content is displayed on replay. The success of a consistent replay experience when the original experience contained a large amount of potential user interactivity (and more importantly, the loading of external resources not initially loaded) might not rely on the level of interactivity able to be re-experienced at replay. The success of this experience is dependent on whether all of the resources needed to properly display the web page on replay were captured at the time of archiving and are loaded when the archive is replayed. 

\begin{figure*}[h!t]
\begin{center}
\subfigure[Live version]{
	\label{fig:gmapsLive}
	\includegraphics[width=0.30\textwidth,natwidth=1016,natheight=940]{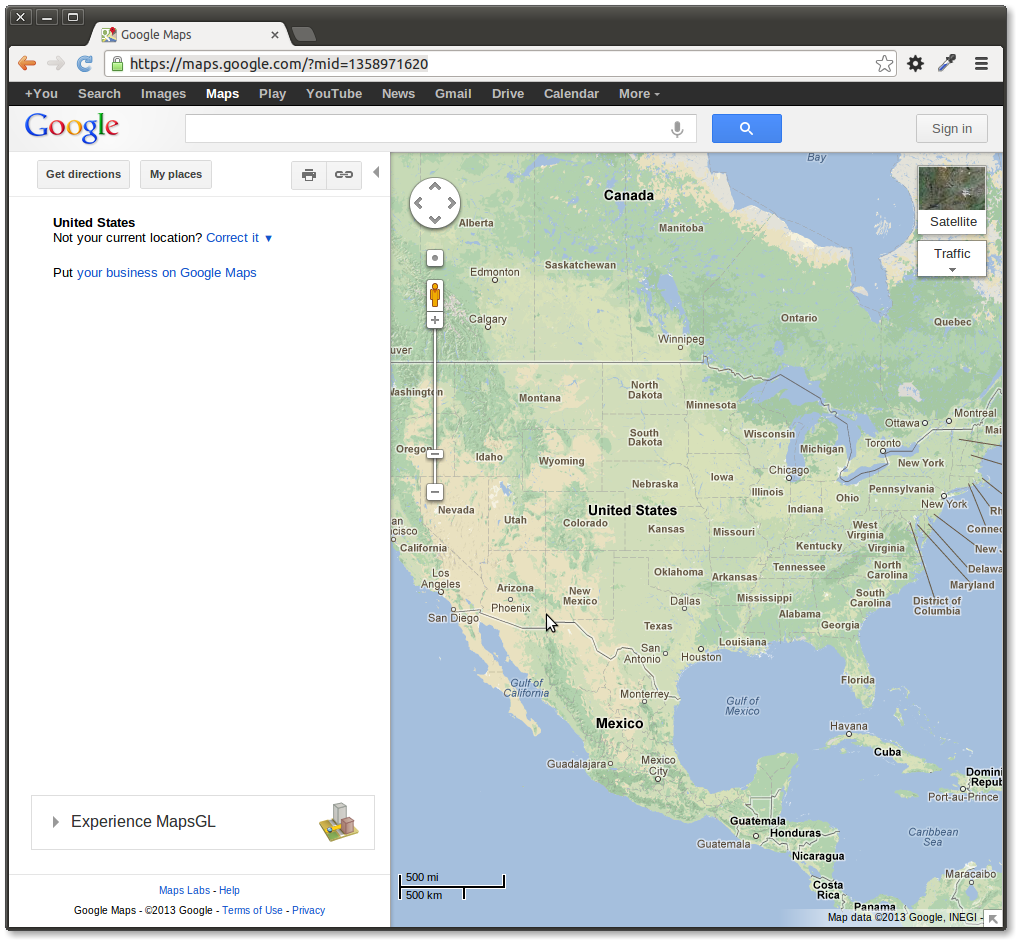}
}
\subfigure[Archived Dec. 1, 2012]{
	\label{fig:gmapsRecent}
	\includegraphics[width=0.30\textwidth,natwidth=1016,natheight=940]{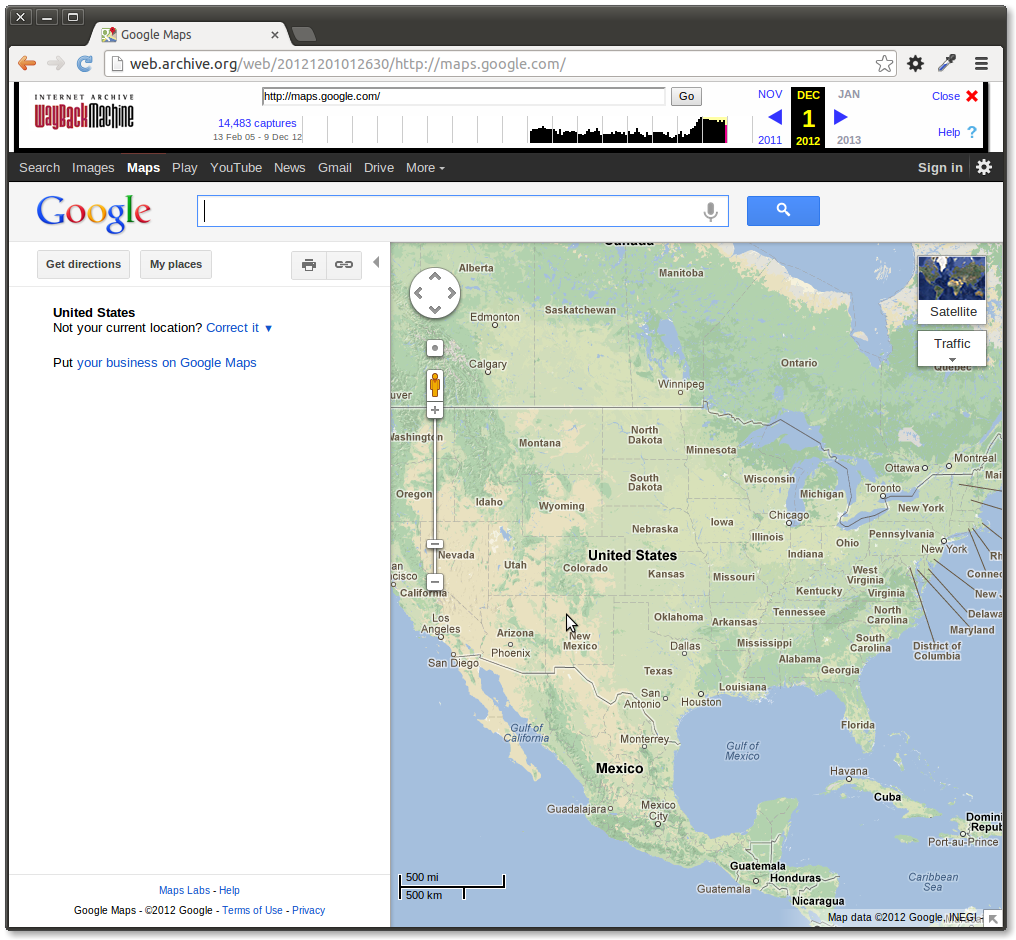}
}
\subfigure[Archived Apr. 30, 2012]{
	\label{fig:gmapsNoJS}
	\includegraphics[width=0.30\textwidth,natwidth=1016,natheight=940]{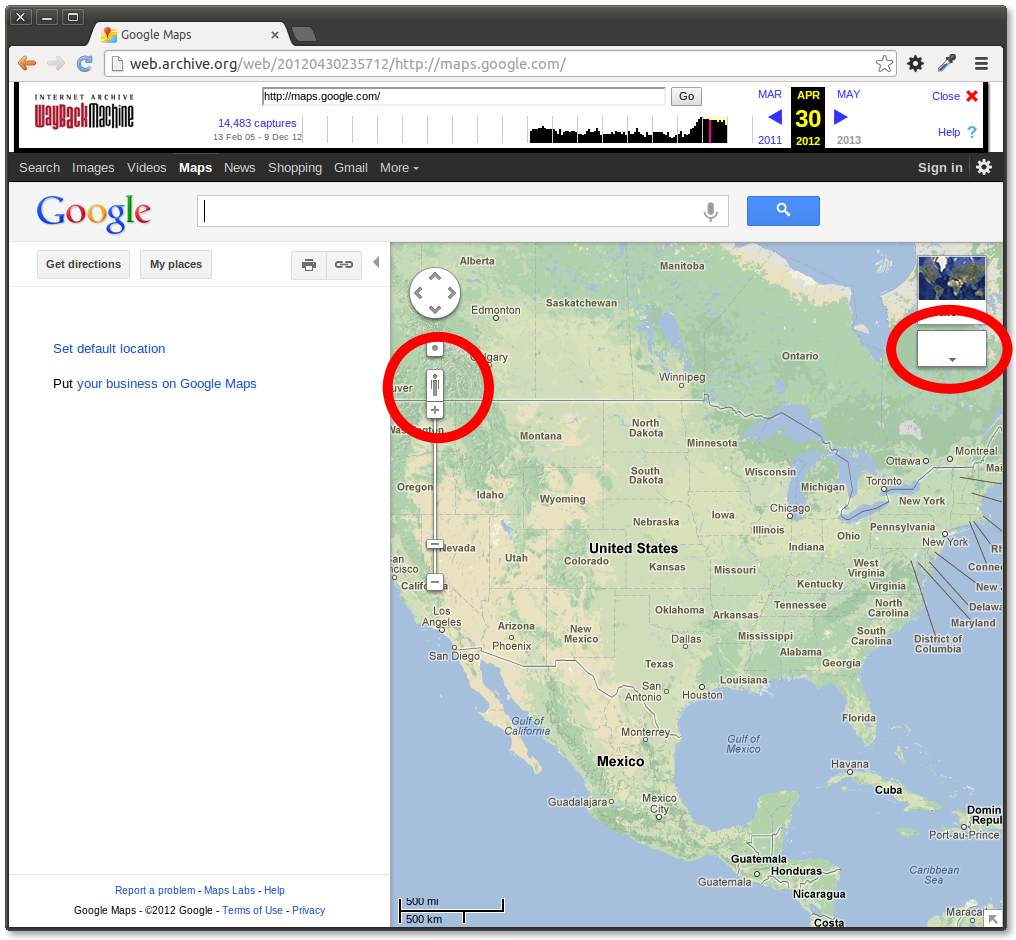}
}

\end{center}	

\caption[test]{Google Maps is a typical example where the lack of completeness of the archive is easy to observe. Figure~\ref{fig:gmapsLive} shows how the web page should look. The map in the middle has interactive UI elements. An archived version of this page (Figure~\ref{fig:gmapsNoJS}) is missing UI elements (circled), and the interaction does not  function. Figure~\ref{fig:gmapsRecent} shows a recently archived (December 1, 2012) version that gives the façade of functionality when, in fact, resources on the live web are being loaded.}
\label{fig:gmaps}
\end{figure*}

The nature of the execution of the archiving procedure is often to blame for not capturing resources loaded at runtime that not only manipulate the Document Object Model (DOM) but also load subsequent representations of resources. These subsequently loaded representations are often captured \cite{mohr2004heritrix} if their location is able to be extracted from the static code by a crawler, but a problem occurs when their loading is latent or triggered by user interaction. An example of this can be seen in the Internet Archive's capture of Google Maps\footnote{\url{http://maps.google.com}} (Figure~\ref{fig:gmaps}). When this archive is replayed, everything that was presented to the user (i.e., the crawler) at time of archiving is displayed. None of the trademark panning or user interface (UI) elements function in a manner similar to the live version (Figure~\ref{fig:gmapsLive}) of the same page. The resources, however, appear to be correctly loaded (Figure~\ref{fig:gmapsNoJS}), though the asynchronous JavaScript calls are never fired because of the broken UI elements. Versions of a page recently archived (Figure~\ref{fig:gmapsRecent}) appear to have all resources required for interactivity, and the page performs as expected, but upon inspection of the URIs, all reference the live web and not the resources at the archive.

Contrasting the completeness of the archive of an interactive website with one from a simpler website that does not contain interactive elements or the loading of external resources via JavaScript further exemplifies that this trait is to blame for archive incompleteness. For example, a page containing only HTML, images, and CSS is likely to be completely represented in the archive when preserved with an archival tool like Heritrix. In this paper we will show that the archivability of a website, given the state of the art of archiving tools, has changed over time with the increased usage of resource-loading JavaScript and the increased accessibility of websites. Further, we will examine the incapability of crawlers and archiving tools in capturing this content and what can be done to remedy their shortcomings and increase the archivability of problematic web pages.

\section{Related Work}
Many components contribute to the archivability of a web page ranging from reliability of the mechanism used to archive to the frequency at which the mechanism is run. Ainsworth et al.\ utilized web directories like DMOZ, Delicious, Bitly and search engine results to determine how much of the web is archived \cite{Ainsworth2011}. McCown et al., in earlier work, developed strategies for resurrecting web pages from the archives but mainly considered those with static resources (including JavaScript) \cite{MCCOWNJCDL07}. McCown also touched on the sources used to recreate lost websites (of particular interest, using Internet Archive's) and the long tail effect on the unlikelihood of domain specific sites to be able to be resurrected using the larger archives as a source \cite{McCown07:WhyWebsitesLost}. Mohr set the basis for the Internet Archive's Heritrix, while introducing incremental crawling into the tool's repertoire of capability \cite{sigurssonheritrix}.

As JavaScript has been the source of many problems in archiving, particularly since the web has become more dynamic, it is useful to note prior attempts relating to JavaScript and archivability. Likarish \cite{DetectingMaliciousJavascript} developed a means of detecting JavaScript with certain facets (namely malicious code via deobfuscation) using Machine Learning techniques and Heritrix. Livshits et al. took on some of the complexities of JavaScript, including attributes only available at runtime and thus normally limited to be experience by the client \cite{kiciman2007ajaxscope,vikram2009ripley,meyerovich2010conscript,livshits2010gulfstream}. Bergman  described the quantity of resources we are unable to index (the ``deep Web'') \cite{bergman2001white}, which has significantly increased with the advent of resources being loaded only at runtime. Ast proposed approaches to capture Ajax content using a conventional web crawler (i.e., one not used for preservation) \cite{ast2008crawler}.

\section{Why JavaScript Makes It Difficult}
A user or script normally browses the web using a user agent. In the case of a user, this is normally a web browser. Initially, web browsers were inconsistent in implementation. This inconsistency eventually was the impetus for creating web standards to remedy the guesswork developers had to do to ensure that the display was as desired. The layout engine is the component of a web browser that is responsible for rendering HTML, the structural portion of a web page. Along with the structure, there is also a stylistic portion (implemented via CSS) and a behavioral portion (implemented in JavaScript) on the client-side. 

As the layout engines of modern browsers evolved, the JavaScript rendering engine lagged behind, causing the behavioral functionality of web pages to perform inconsistently among users. This was particularly noticeable when Ajax-based websites became common. To simplify the process of obtaining the data quickly and without worry about behavior, many crawlers and scrapers do not include a JavaScript rendering engine, opting to only grab the HTML and any embedded resources and rely on the user's layout engine to render the fetched data at a later date. This is problematic in that some resources' location on the web might be built at runtime or included in the page because of JavaScript DOM manipulation. In the case of a crawler, this might be negligible, as the resource will still be hot-linked and thus included when the web page is ``replayed''. For crawlers intended for preservation, however, the archive must be self-contained and thus, for a complete archive, these resources must be captured and their locations rewritten to be accessible at time of replay.

Early versions of Heritrix had limited support for JavaScript. The crawler's parsing engine attempted to detect URIs in scripts, fetch the scripts and recursively repeat this process with the intention of ensuring maximum coverage of the archive creation process. Recently, Heritrix was rewritten to be built on top of PhantomJS\footnote{http://phantomjs.org/}, a headless WebKit (the layout engine used by Google Chrome, Apple Safari, etc.) with JavaScript support, which greatly increases the potential for accurate JavaScript processing and the likelihood that all resources required to replay a web page are captured by a crawler.

\section{Archivability of Sites in Respect to Type}

\label{sec:type}

In one of our preliminary studies we found that web pages of links shared over Twiter were less archivable than those selected to be preserved by the collection-based Archive-It service. Interestingly, many of the websites on Archive-It are governmental.  Many of those on Twitter are commercially-based. Unlike commercial websites, governmental websites are mandated to conform to web accessibility standards for content.  From this, it can be extrapolated that the more accessible a website is, the more archivable it is. Further, certain features of JavaScript (e.g., the reliance of it being enabled to show content) may make a page generally less accessible.

The markup (HTML) of a web page is rarely a hindrance to a website being captured. 
The difference in interpretation of markup among various users, for instance, does not make the code more or less accessible. Semantic markup, as encouraged by Section 508, WAI specifications, and other organizations that advocate accessible web development practices does affect how end-users see the content. Even if the content displayed is hidden from view, it is still likely present and thus preserved, making variance in markup replay a moot point.

In contrast to markup, behavior (usually JavaScript) can contend what content resides in the markup. If certain behaviors are not invoked, certain content may never make it to the markup, thus compromising the degree at which the content that should be archived is archived. 

\section{Experimental Setup}
\label{sec:expSetup}
We used the Memento Framework \cite{nelsonmemento} to query mementos with a reliable datetime association. Using a URI convention and the HTTP Accept-Datetime header, archives can be queried for the approximate time desired and the closest result will be returned.

\subsection{Gauging Accessibility}
\label{sec:accessibility}
While Section 508 gives suggestions on how websites should comply to be considered accessible, other organizations (namely the World Wide Web Consortium (W3C) through the Web Content Accessibility Guidelines (WCAG)) give concrete ways for developers to evaluate their creations to make them more accessible \cite{wcag}. Hackett et al. go in-depth on the accessibility of archives on a pre-Ajax corpus (from 2002) and enumerate very specific features that make a page accessible in the legal sense. Much of this information is beyond the scope of this study. 
Other efforts were made following the document's release on a metric to evaluate web accessibility \cite{Parmanto}, but also fail to consider an asynchronous web.

As stated in Section~\ref{sec:type}, the markup of a web page is not problematic for crawlers to capture. To understand the role that JavaScript plays in hindering comprehensive web archiving by crawlers, it is useful to examine the WCAG Principles of Accessibility \cite{wcag} and remark on where issues would occur for a crawler. JavaScript, specifically, affects a page's accessibility by hiding information from view (perceivability), only working if all components of the script are present (operability), and frequently playing a critical role in a page being useable (robustness).

\begin{figure}[h]
\begin{algorithmic}[1]\scriptsize
\Function{GetMemsWithYearInterval}{mementoURIs} 
 \State $M\gets mementoURIs[1]$ \Comment{Get First Memento}
 \State $lastDate\gets extractDate(M)$
 \State $lastDateTest\gets lastDate$
 \For{$m = 2 \to length(mementoURIs)$} 
	\State $testingDate\gets extractDate(mementoURIs[m])$
  	\State $testingDate = extractDate(mementoURIs[m])$ 
  	\If{$lastDate + oneYear\le testingDate$}
      \If{$| lastDate - testingDate + oneYear | \ge$\\
             \hspace{1.8cm}$| lastDateTested - lastDate + oneYear |$ }
        \State $lastDate\gets mementoURIs[m-1]	$
      \Else
        \State $lastDate\gets mementoURIs[m]$
       
      \EndIf
      \State push(M,mementoURIs[m])
    \Else 
       \State $lastDateTested\gets testingDate$	      
    \EndIf
 \EndFor
 
 \State \Return $M$
\EndFunction
\end{algorithmic}
\caption{Pseudocode to get mementos from a timegate at a temporal distance of one year per memento or as close as possible with the first memento as the pivot.}
\label{fig:oneYearSpreadCode}
\end{figure}

\subsection{Fetching Data}

The initial experiment was to test the archivability of web sites whose presence in the archive has persisted over a long period of time. These can be described as the ``stubby head'' juxtaposed to McCown's ``long tail'' of sites that are preserved. Alexa\footnote{\url{http://www.alexa.com}} has gathered the traffic of many of these sites and ranked them in descending order. This ranking currently exists as Alexa's Top 500 Global Sites\footnote{\url{http://www.alexa.com/topsites}}. 
We first attempted an approach at gathering data by querying the archives (namely, Internet Archive's Wayback) for past Top lists\footnote{\url{http://web.archive.org/web/20090315000000*/http://www.alexa.com/topsites}} but found that the location of this list was inconsistent to the present one and some of the sites in past top 10 lists remained present in the current list. We used a simple scraping scheme to grab the paginated list but that turned up pornographic sites by the third page on the 2012 list so we kept it to the top few sites to remain representative, unbiased, and to reduce the likelihood of including sites without longevity.

\begin{figure*}[h]
\centering
\begin{lstlisting}
<http://api.wayback.archive.org/list/timebundle/http://cnn.com>; rel="timebundle",
<http://cnn.com>; rel="original",
<http://api.wayback.archive.org/list/timemap/link/http://cnn.com>; rel="timemap"; type="application/link-format",
<http://api.wayback.archive.org/list/timegate/http://cnn.com>; rel="timegate",
<http://api.wayback.archive.org/memento/20000620180259/http://cnn.com/>; rel="first memento"; datetime="Tue, 20 Jun 2000 18:02:59 GMT",
<http://api.wayback.archive.org/memento/20000621011731/http://cnn.com/>; rel="memento"; datetime="Wed, 21 Jun 2000 01:17:31 GMT",
<http://api.wayback.archive.org/memento/20000621140928/http://cnn.com/>; rel="memento"; datetime="Wed, 21 Jun 2000 14:09:28 GMT",
...
<http://api.wayback.archive.org/memento/20061227222050/http://www.cnn.com>; rel="memento"; datetime="Wed, 27 Dec 2006 22:20:50 GMT",
<http://api.wayback.archive.org/memento/20061227222134/http://www.cnn.com/>; rel="memento"; datetime="Wed, 27 Dec 2006 22:21:34 GMT",
<http://api.wayback.archive.org/memento/20061228024612/http://www.cnn.com/>; rel="memento"; datetime="Thu, 28 Dec 2006 02:46:12 GMT",
...
<http://api.wayback.archive.org/memento/20121209174923/http://www.cnn.com/>; rel="memento"; datetime="Sun, 09 Dec 2012 17:49:23 GMT",
<http://api.wayback.archive.org/memento/20121209174944/http://www.cnn.com/>; rel="memento"; datetime="Sun, 09 Dec 2012 17:49:44 GMT",
<http://api.wayback.archive.org/memento/20121209201112/http://www.cnn.com/>; rel="last memento"; datetime="Sun, 09 Dec 2012 20:11:12 GMT"
\end{lstlisting}
\caption{A sample abbreviated (for space) TimeMap for cnn.com}
\label{fig:timemap}
\end{figure*}

For each of these web sites, the TimeMap was acquired from Internet Archive using the URI convention ``\url{http://api.wayback.archive.org/list/timemap/link/}\textit{URI}'' (where \textit{URI} is the Fully Qualified Domain Name of the target site) to produce output like Figure~\ref{fig:timemap}. From this TimeMap we chose mementos with a one year spread. 

The URI of each memento was passed to a PhantomJS script, and the HTTP codes of each resource as well as a snapshot were taken. 
A memento\footnote{\url{http://api.wayback.archive.org/memento/20110731003335/http://google.com}} for \url{google.com} with the Memento-Datetime 20110731003335, for example, produces a line break delimited list of the subsequent HTTP codes and respective URIs dereferenced to assemble the page. Here, we noticed that subsequent requests for the resources yielded resources from the live web. We tailored the PhantomJS script to rewrite the URIs to hit the Wayback Machine instead\footnote{e.g., \url{http://web.archive.org/web/20110731003335/http://google.com}}. This produces an identical display (part of the Wayback UI was programatically hidden) but with resource requests that access archived content. 
The last step was repeated but with PhantomJS sent the directive to capture the page with JavaScript off.

From the top 10 websites on Alexa for 2012, some websites had a robots.txt restriction. The number of mementos obtained by using the code in Figure~\ref{fig:oneYearSpreadCode} and applying the URI transformation produces the following quantity of mementos, ordered corresponding to Alexa's 2012 ranking (Table~\ref{tab:alexa}).

\begin{table}
\centering
\small
\begin{tabular}{|c|c|c|}
\hline
Alexa Rank & Web Site Name & Available Mementos \\ \hline
1 & Facebook.com & \specialcell{no mementos robots.txt exclusion}\\ \hline
2 & Google.com & \specialcell{15 mementos 1998 to 2012}\\ \hline
3 & YouTube.com & \specialcell{7 mementos 2006 to 2012}\\ \hline
4 & Yahoo.com & \specialcell{16 mementos 1997 to 2012}\\ \hline
5 & Baidu.com & \specialcell{no mementos robots.txt exclusion}\\ \hline
6 & Wikipedia.org & \specialcell{12 mementos 2001 to 2012}\\ \hline
7 & Live.com & \specialcell{15 mementos 1999 to 2012}\\ \hline
8 & Amazon.com & \specialcell{14 mementos 1999 to 2012}\\ \hline
9 & QQ.com & \specialcell{15 mementos 1998 to 2012}\\ \hline
10 & Twitter.com & \specialcell{no mementos robots.txt exclusion}\\ \hline
\end{tabular}
\begin{center}
\caption{Alexa's 2012 Top 10 websites and available mementos.}
\end{center}
\label{tab:alexa}
\end{table}
That the content of some of these websites is not preserved (namely Facebook and Twitter) by institutions has been addressed by multiple parties \cite{1555440,kelly-jcdl12,loctweets,bass2012getting,meyerresearcher}. 
All of these websites may not exhibit traits of un-archivability, as previously imagined. The root domain, in this case, may not be representative of the extent at which a web site (contrasted to web page) utilized un-archivable practices. One particular site that has succumbed to the effects of unarchivability due partially to both its longevity and publishing medium is YouTube. Crook \cite{crook2009web} went into detail about the issues in preserving multimedia resources on the web, and Prellwitz documented how quickly this multimedia degrades \cite{Prellwitz}, so highlighting this website for analysis would be useful in remedying one of the many reasons that it is not sufficiently preserved.

\begin{figure}[h]
\centerline{
\subfigure[]{
	\label{fig:youtubeWithJS}
	\includegraphics[width=0.4\textwidth,natwidth=883,natheight=604]{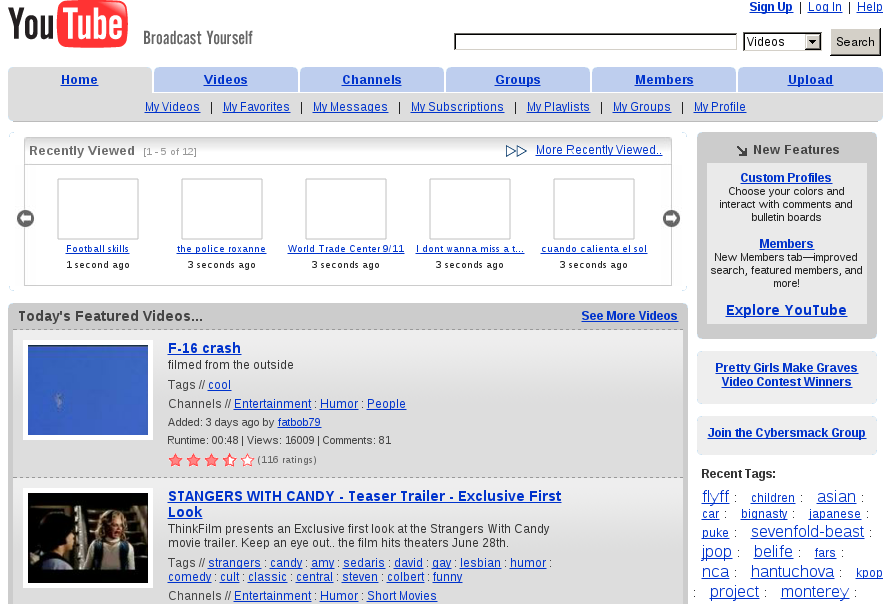}
}
\subfigure[]{
	\label{fig:youtubeWithoutJS}
	\includegraphics[width=0.4\textwidth,natwidth=883,natheight=604]{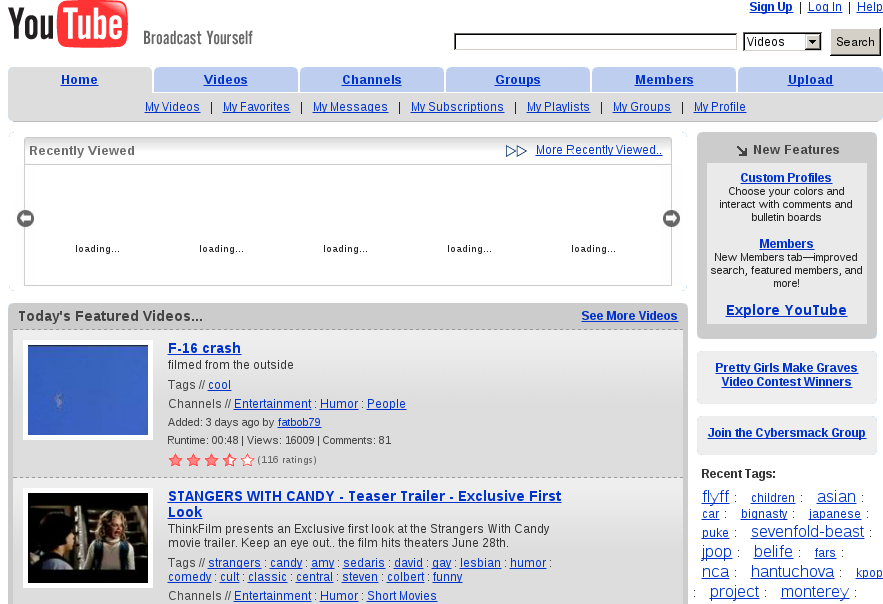}
}
}
\caption[test]{A YouTube memento from 2006 shows a subtle distinction in display when JavaScript is enabled (Figure~\ref{fig:youtubeWithJS}) and disabled (Figure~\ref{fig:youtubeWithoutJS}) at the time of capture. The Ajax spinner (above each ``loading'' message in Figure~\ref{fig:youtubeWithoutJS}) is never replaced with content, which would be done were JavaScript enabled on capture. When it was enabled, the script that gathers the resources to display (blank squares in the same section of the site in Figure~\ref{fig:youtubeWithJS}) is unable to fetch the resources it needs in the context of the archive. The URIs of each of these resources (the image source) is present as an attribute of the DOM element but because it is generated post load, the crawler never fetches the resource for preservation.}
\label{fig:youtube}
\end{figure}

Using the procedure described earlier in this section, we captured screen shots and HTTP requests for one memento per year of YouTube.com. While there have been efforts in attempting to capture the multimedia on this site in a reliable way (e.g., TubeKit\cite{shah2008tubekit}), our concern is less about executing a focused crawl and more on analyzing the results of what has been done in the past.  The simpler case here of lack of archivability is observable from the homepage. In each of the cases of capturing a screen shot (Figure~\ref{fig:youtube}) of the memento with and without JavaScript, there is variance on the ``Recently Viewed'' section of the website. This part of the website is Ajax-driven, i.e., after the page has loaded, the content is fetched. A crawler could retain the JavaScript that fetches the resources and attempt to grab a copy of the resources contained within and loaded at runtime but this particular script takes a moment post-load to load and display the images that represent links to videos. This is better explained by Figure~\ref{fig:youtubeWithJS}, which is representative of the memento with JavaScript enabled. The content necessary to display this section was preserved due to its reliance on runtime execution. Figure~\ref{fig:youtubeWithoutJS} shows the same memento fetched with JavaScript off. The place-holder Ajax ``spinner'' demonstrates that the JavaScript to overwrite the DOM elements is present in the archive and executable but the resources needed to fully build this web page do not exist in the archive. 

\begin{figure}[h]
\centering
\subfigure[]{
	\label{fig:youtube2011screenshot}
	\includegraphics[width=0.4\textwidth,natwidth=1156,natheight=728]{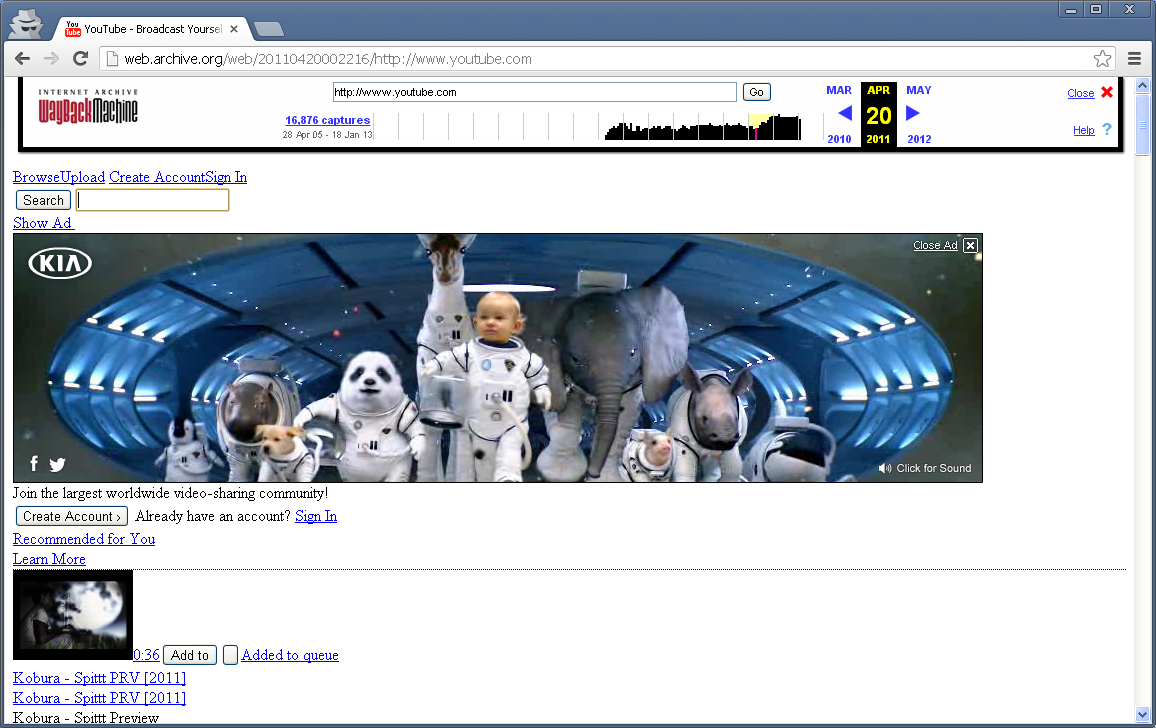}
}
\subfigure[]{
	\label{fig:youtube2011errors}
\begin{lstlisting}
GET http://web.archive.org/web/20121208145112cs_/http://s.ytimg.com/yt/cssbin/www-core-vfl_OJqFG.css 404 (Not Found) www.youtube.com:15
GET http://web.archive.org/web/20121208145115js_/http://s.ytimg.com/yt/jsbin/www-core-vfl8PDcRe.js 404 (Not Found) www.youtube.com:45
Uncaught TypeError: Object #<Object> has no method 'setConfig' www.youtube.com:56
Uncaught TypeError: Cannot read property 'home' of undefined www.youtube.com:76
Uncaught TypeError: Cannot read property 'ajax' of undefined www.youtube.com:86
Uncaught TypeError: Object #<Object> has no method 'setConfig' www.youtube.com:101
Uncaught ReferenceError: _gel is not defined www.youtube.com:1784
Uncaught TypeError: Object #<Object> has no method 'setConfig' www.youtube.com:1929
Uncaught TypeError: Cannot read property 'home' of undefined www.youtube.com:524
GET http://web.archive.org/web/20130101024721im_/http://i2.ytimg.com/vi/1f7neSzDqvc/default.jpg 404 (Not Found) 
\end{lstlisting}
}
\caption{The 2011 capture of this YouTube.com memento demonstrates the causal chain that occurs when a resource is not captured.}
\label{fig:youtube2011}

\end{figure}

Contrast this to five years later (2011) when a redesign of YouTube that is heavily reliant on Ajax fails. When loading this memento\footnote{\url{http://api.wayback.archive.org/memento/20110420002216/http://youtube.com}} into Wayback via a web browser, the JavaScript errors in Figure~\ref{fig:youtube2011errors} appear in the console. This memento (Figure~\ref{fig:youtube2011screenshot}) exhibits leakage \cite{zombies}.

The lack of aesthetic of the 2011 YouTube memento is a result of the CSS files (first line of Figure~\ref{fig:youtube2011errors}) returning an HTTP 302 with the final URI resulting in a 404, as evidence in the log file that accompanies the Figure~\ref{fig:youtube2011screenshot} screenshot during the annual memento collection process. By examining the JavaScript log, we noted that a causal chain prevented subsequent resources from being fetched. JavaScript is fairly resilient to runtime errors and oftentimes will continue executing so long as a resource dependency is not hit\footnote{This is by design of the interpreted language but appears to go against the fast-fail software philosophy.}. The progressive increase of Ajax on YouTube over time has caused a longer chain of failures than the 2006 example. Testing this same procedure on a website that persisted from before Ajax existed until today yet chose to rely on it at one time would test whether its inclusion greatly reduced the archivability.

\section{A Reinforcing Case}

\begin{figure*}[h]
\begin{center}
\centering
\subfigure[1996]{
	\label{fig:nasa1996}
	\includegraphics[width=0.08\linewidth,natwidth=818,natheight=1915]{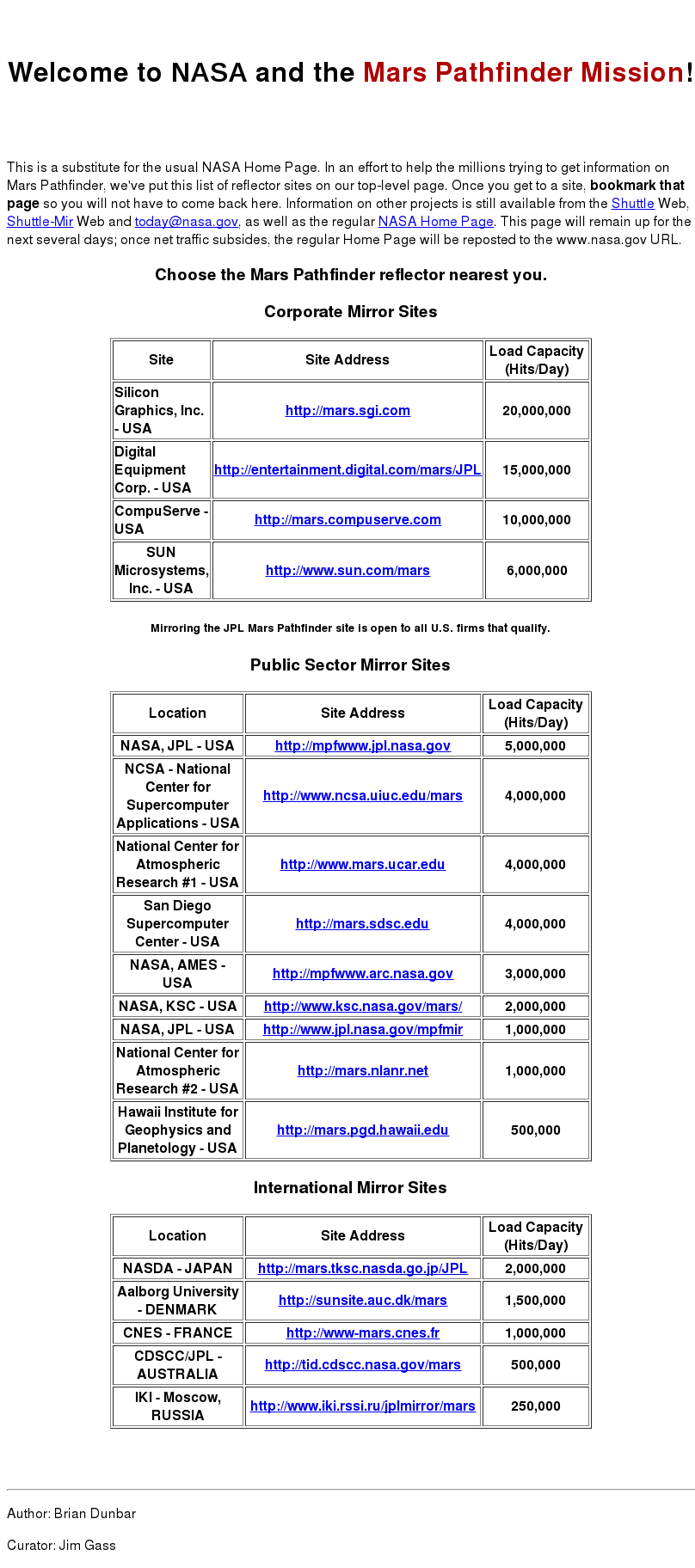}
}
\subfigure[1997]{
	\label{fig:nasa1997}
	\includegraphics[width=0.08\linewidth,natwidth=818,natheight=1915]{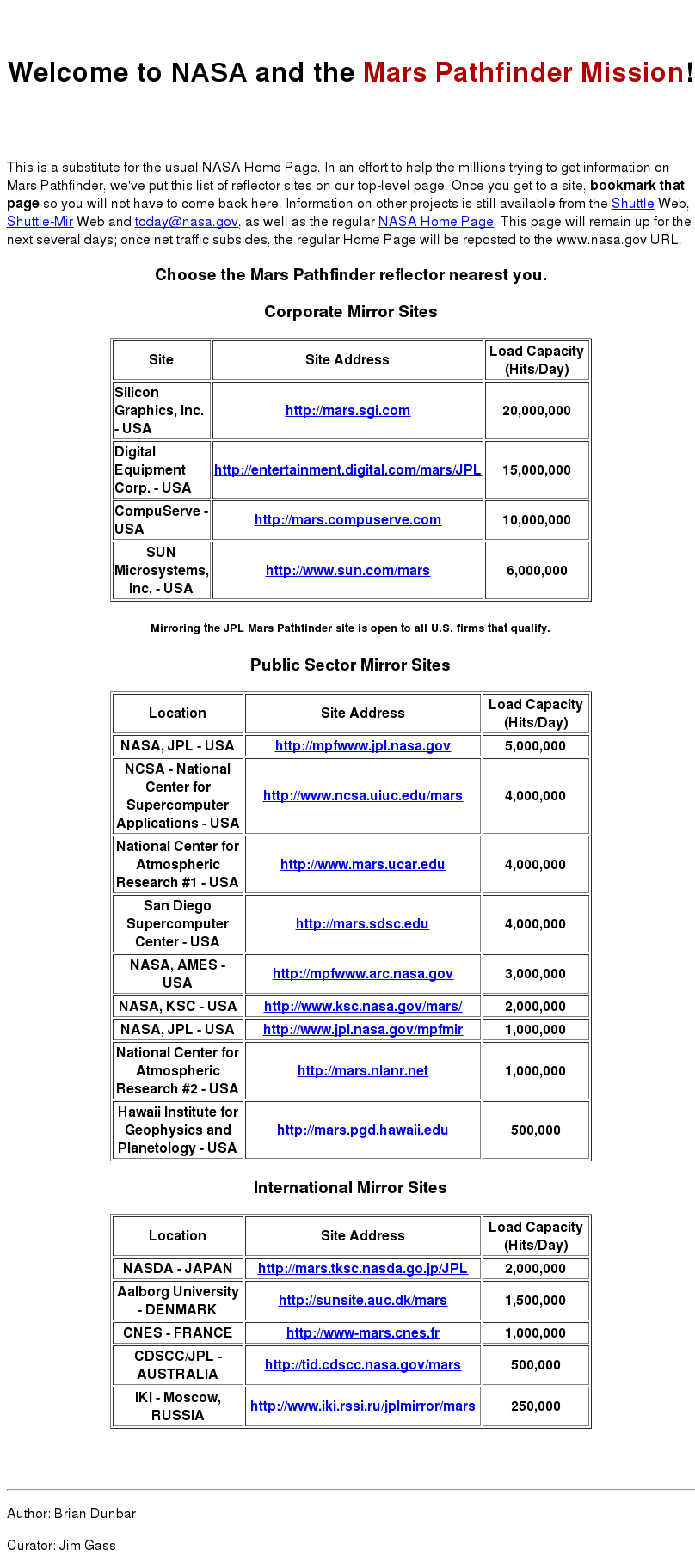}
}
\subfigure[1998]{
	\label{fig:nasa1998}
	\includegraphics[width=0.08\linewidth,natwidth=818,natheight=1915]{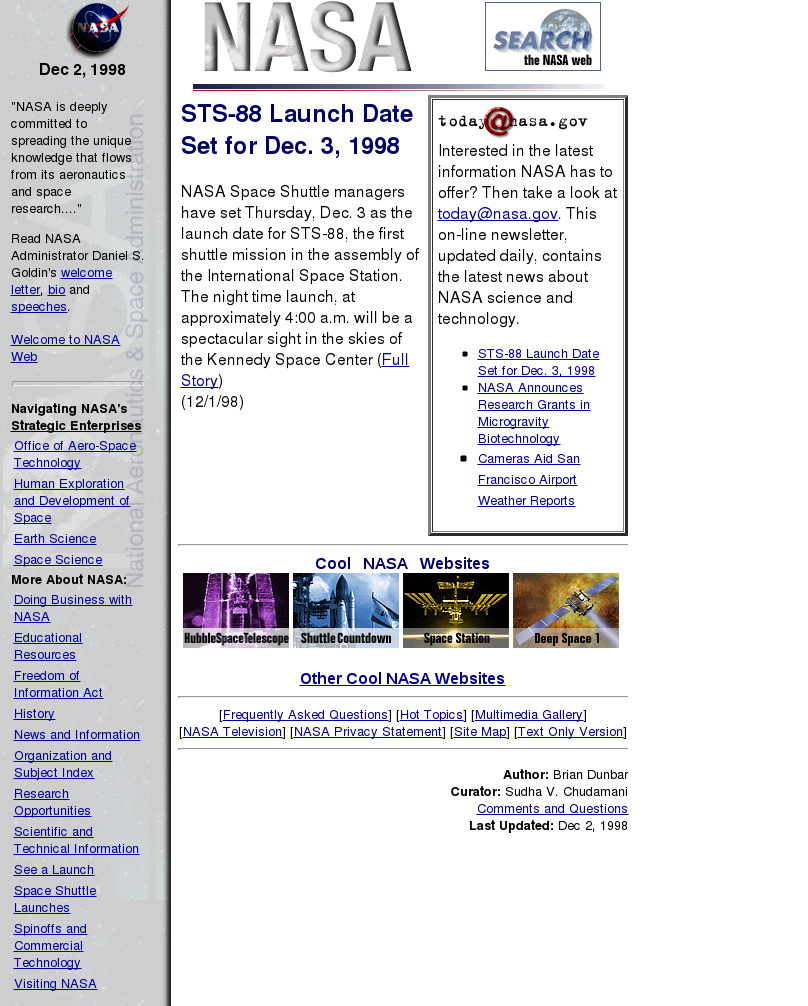}
}
\subfigure[1999]{
	\label{fig:nasa1999}
	\includegraphics[width=0.08\linewidth,natwidth=818,natheight=1915]{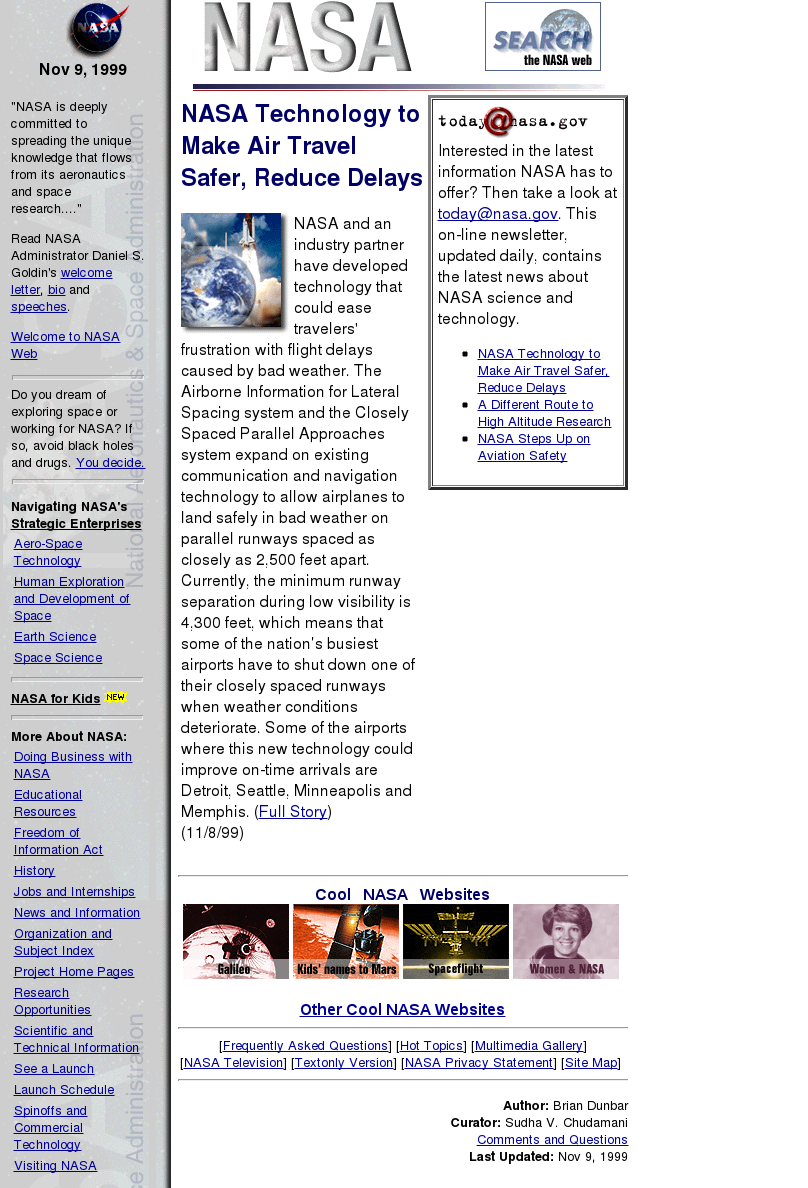}
}
\subfigure[2000]{
	\label{fig:nasa2000}
	\includegraphics[width=0.08\linewidth,natwidth=818,natheight=1915]{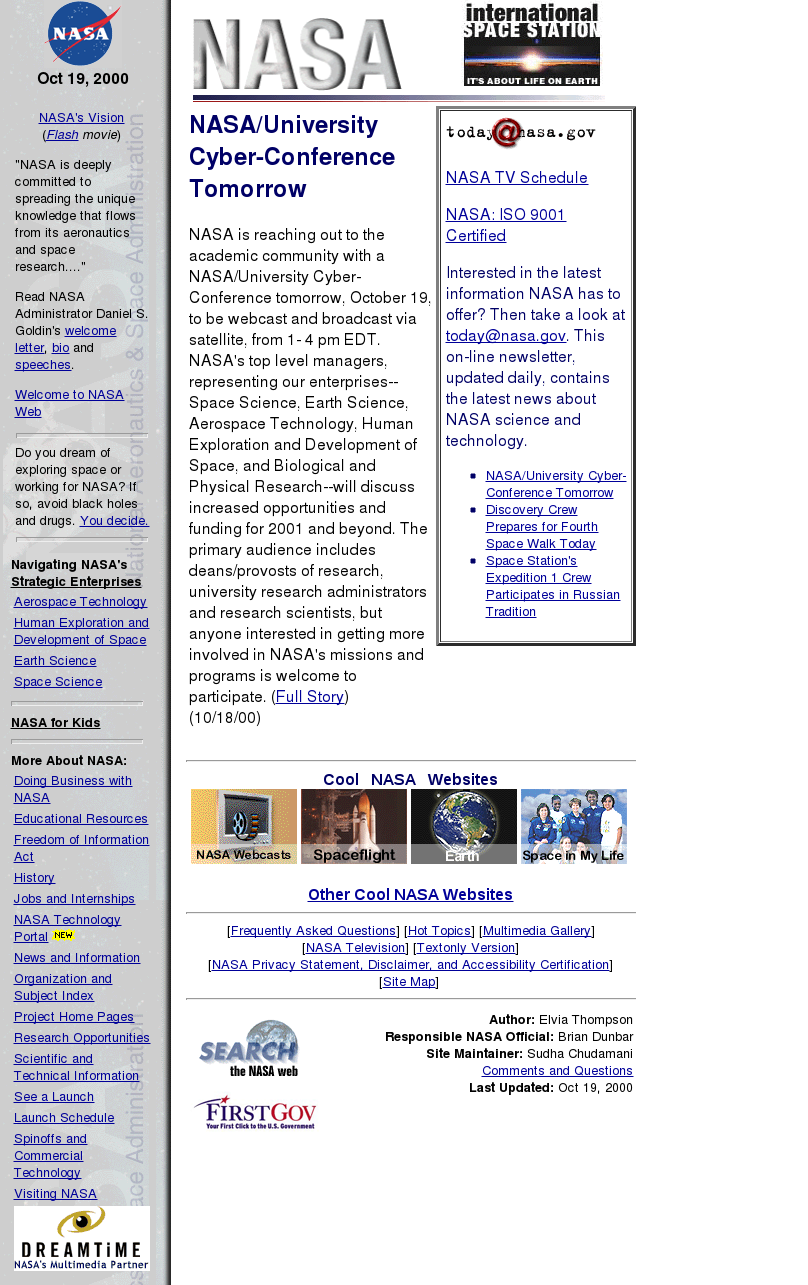}
}
\subfigure[2001]{
	\label{fig:nasa2001}
	\includegraphics[width=0.08\linewidth,natwidth=818,natheight=1915]{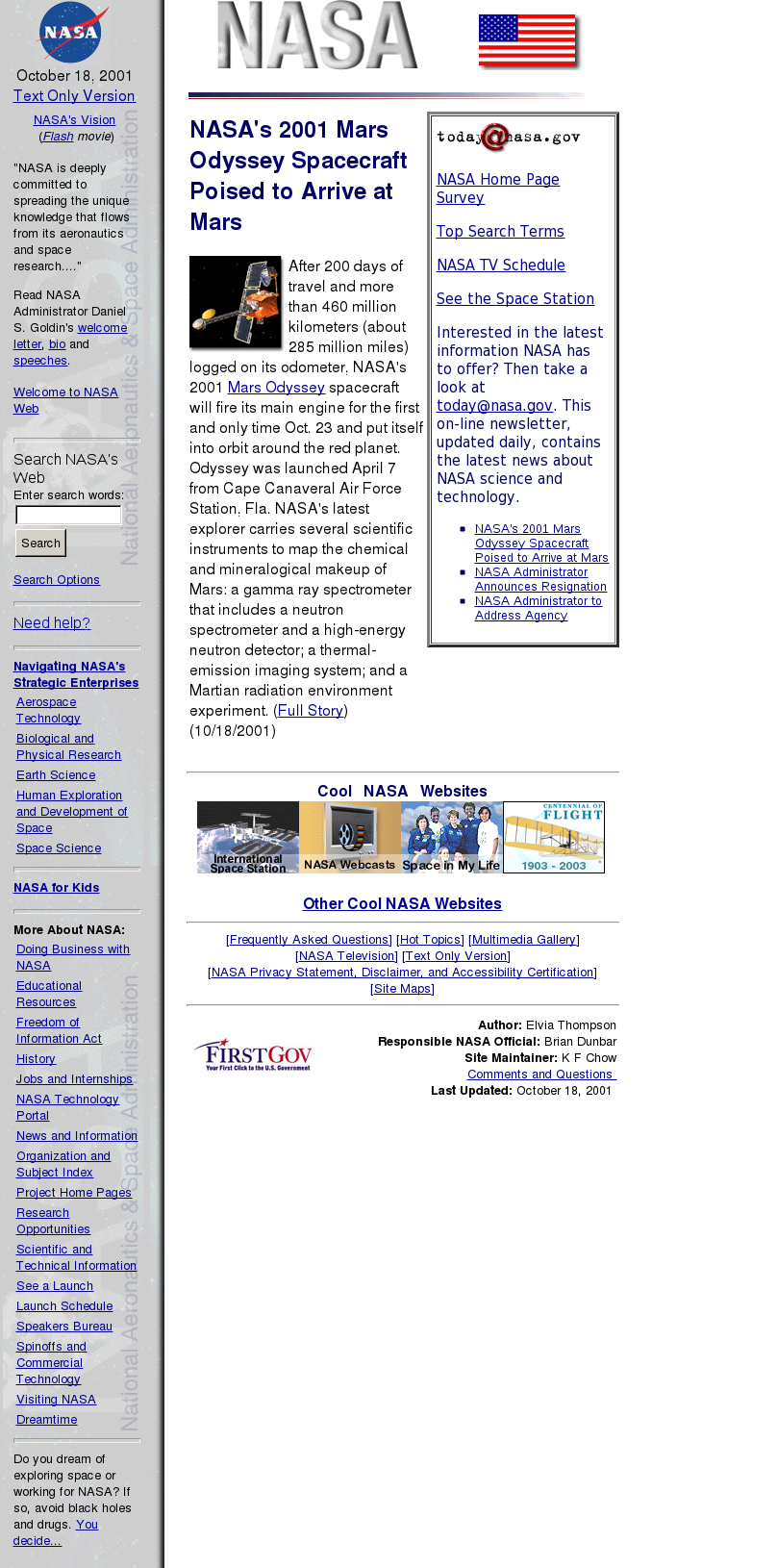}
}
\subfigure[2002]{
	\label{fig:nasa2002}
	\includegraphics[width=0.08\linewidth,natwidth=818,natheight=1915]{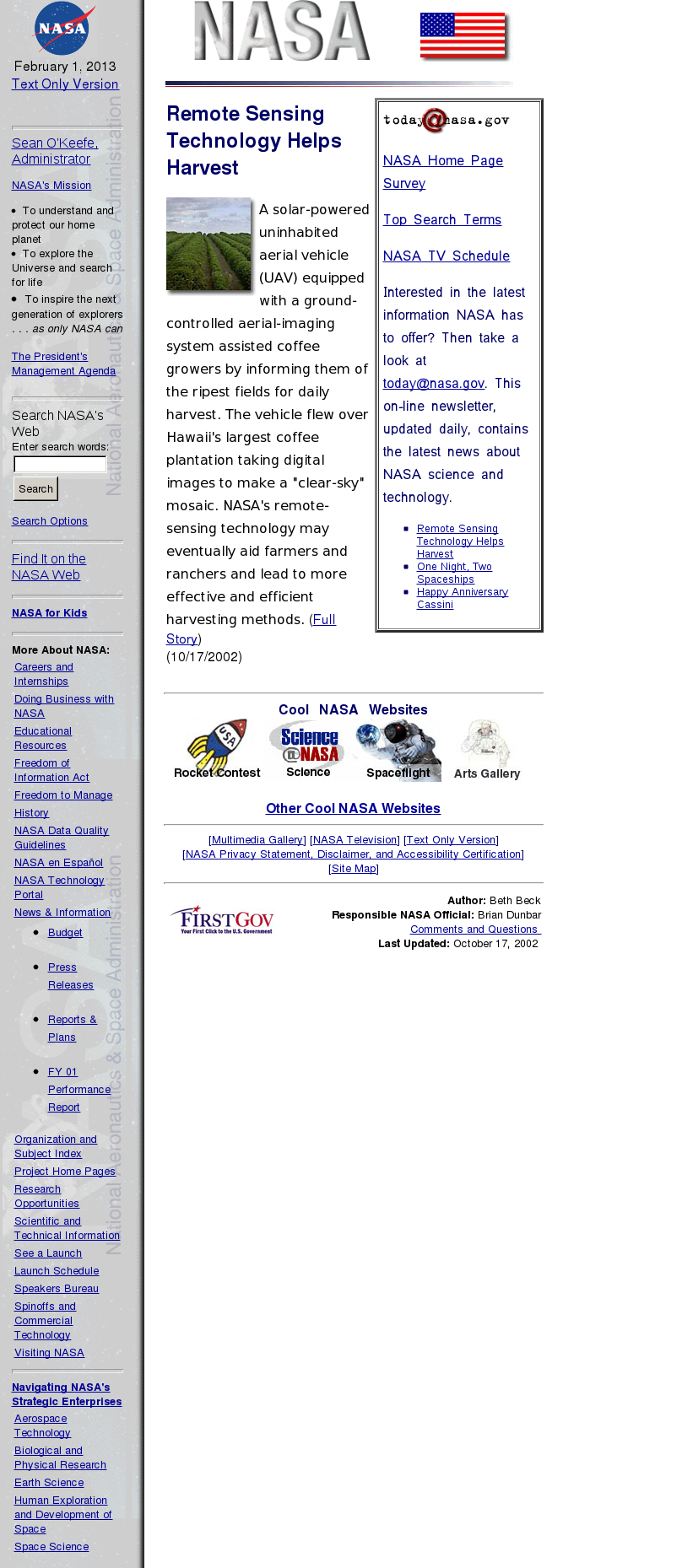}
}
\subfigure[2003]{
	\label{fig:nasa2003}
	\includegraphics[width=0.08\linewidth,natwidth=818,natheight=1915]{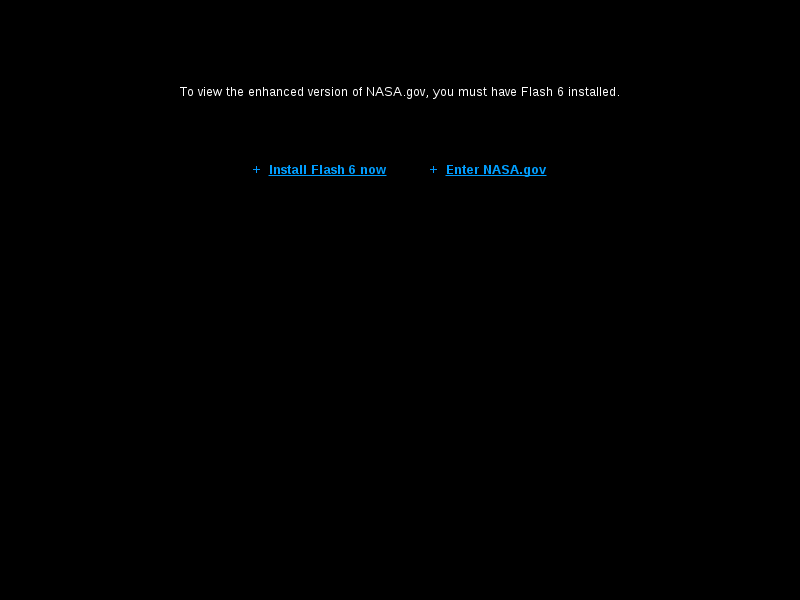}
}
\subfigure[2004]{
	\label{fig:nasa2004}
	\includegraphics[width=0.08\linewidth,natwidth=818,natheight=1915]{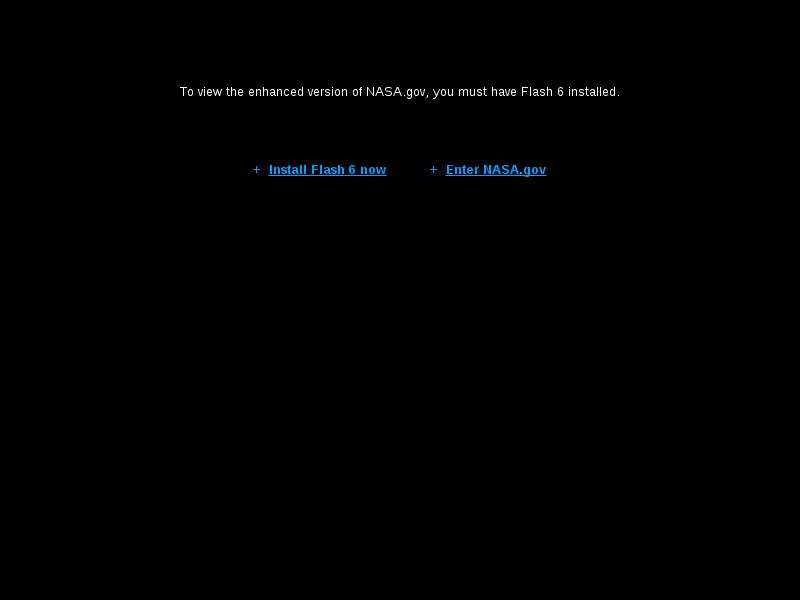}
}
\subfigure[2005]{
	\label{fig:nasa2005}
	\includegraphics[width=0.08\linewidth,natwidth=818,natheight=1915]{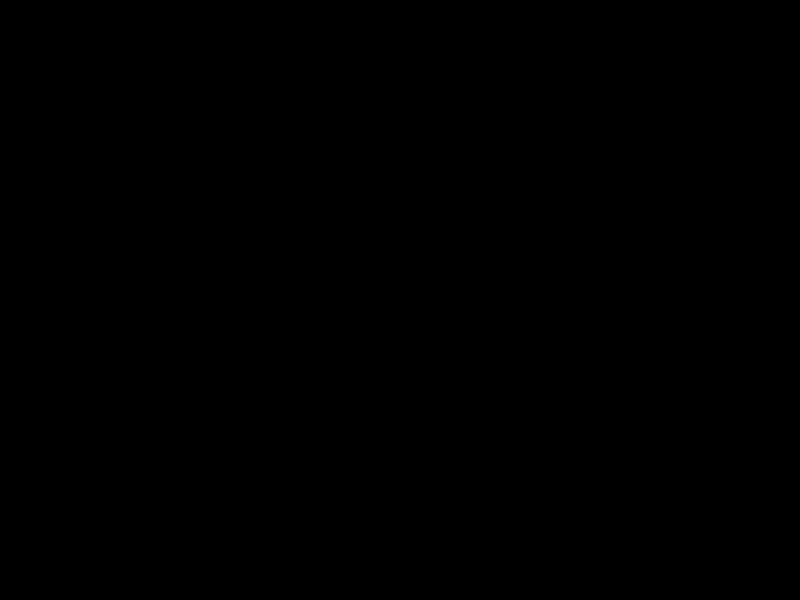}
}
\subfigure[2006]{
	\label{fig:nasa2006}
	\includegraphics[width=0.08\linewidth,natwidth=818,natheight=1915]{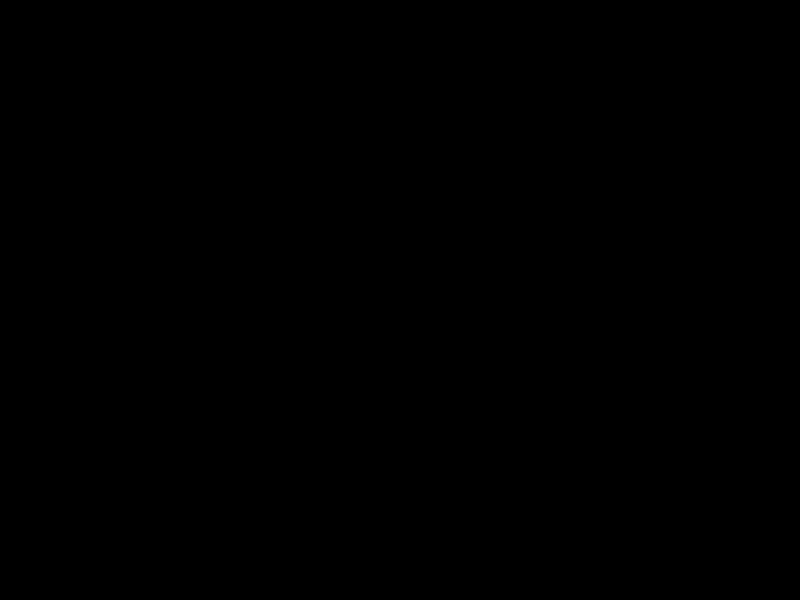}
}
\subfigure[2007]{
	\label{fig:nasa2007}
	\includegraphics[width=0.08\linewidth,natwidth=818,natheight=1915]{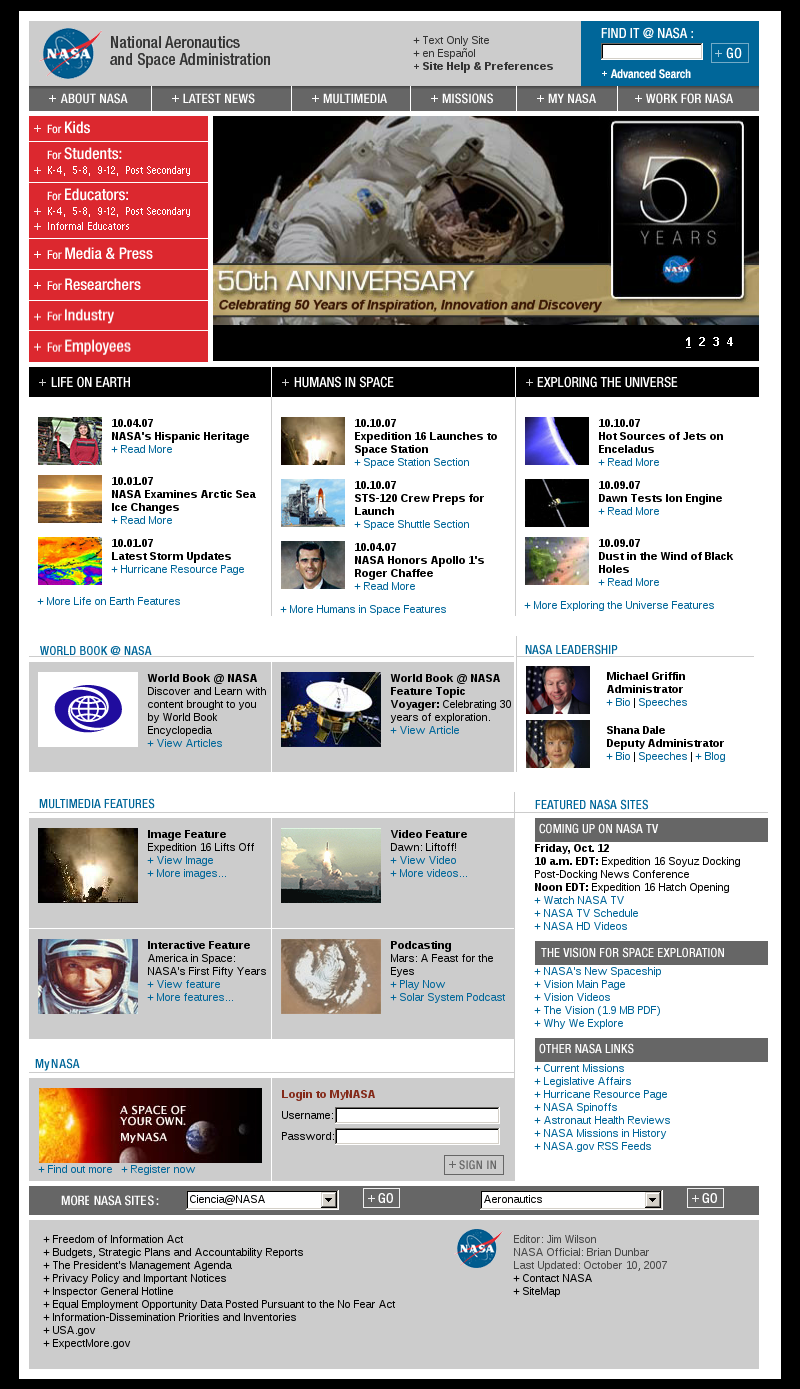}
}
\subfigure[2008]{
	\label{fig:nasa2008}
	\includegraphics[width=0.10\linewidth,natwidth=818,natheight=1915]{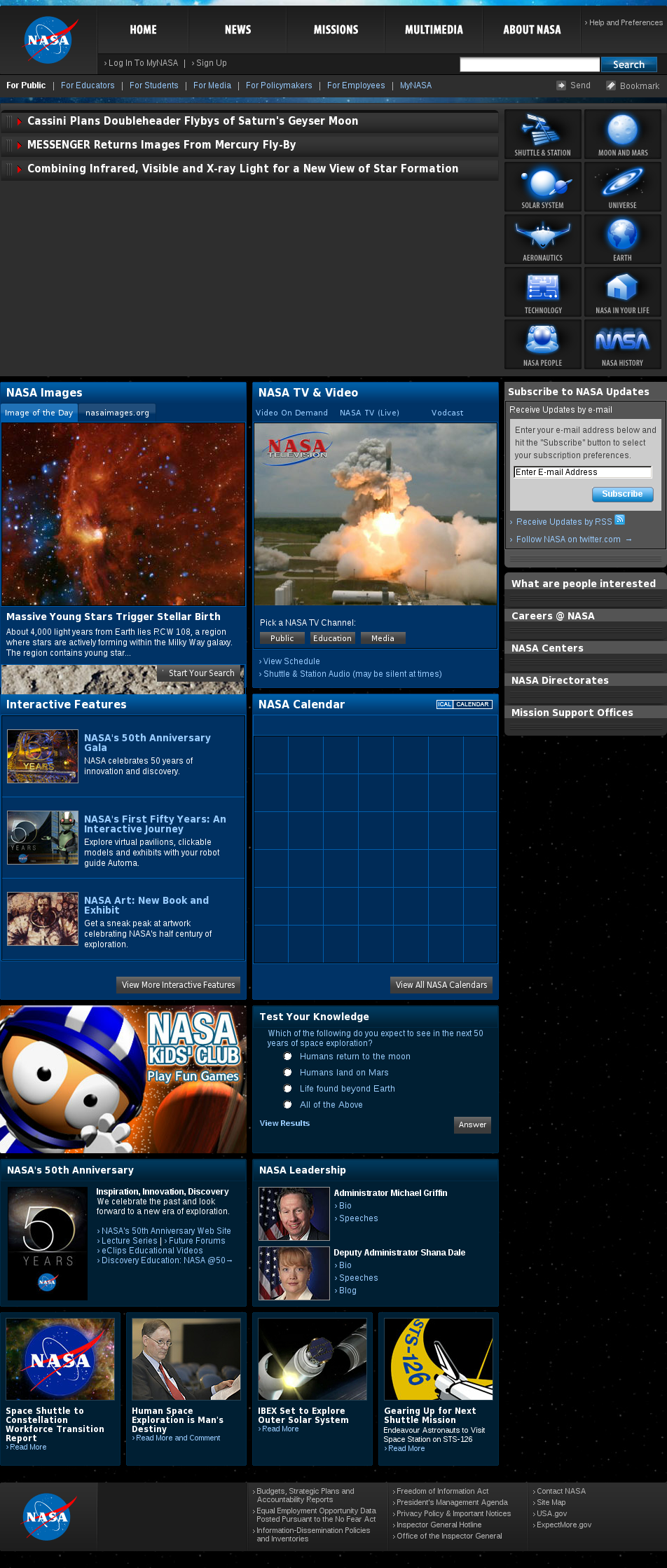}
}
\subfigure[2009]{
	\label{fig:nasa2009}
	\includegraphics[width=0.08\linewidth,natwidth=818,natheight=1915]{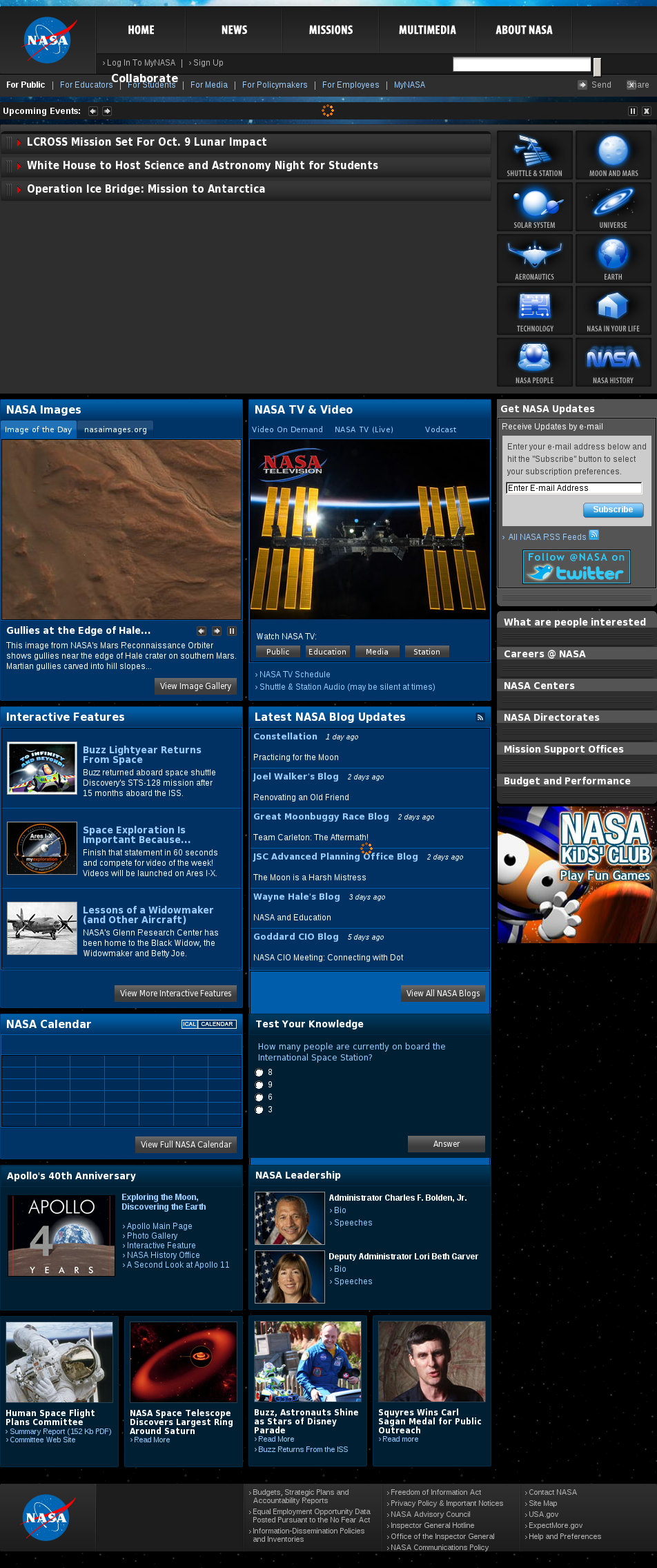}
}
\subfigure[2010]{
	\label{fig:nasa2010}
	\includegraphics[width=0.08\linewidth,natwidth=818,natheight=1915]{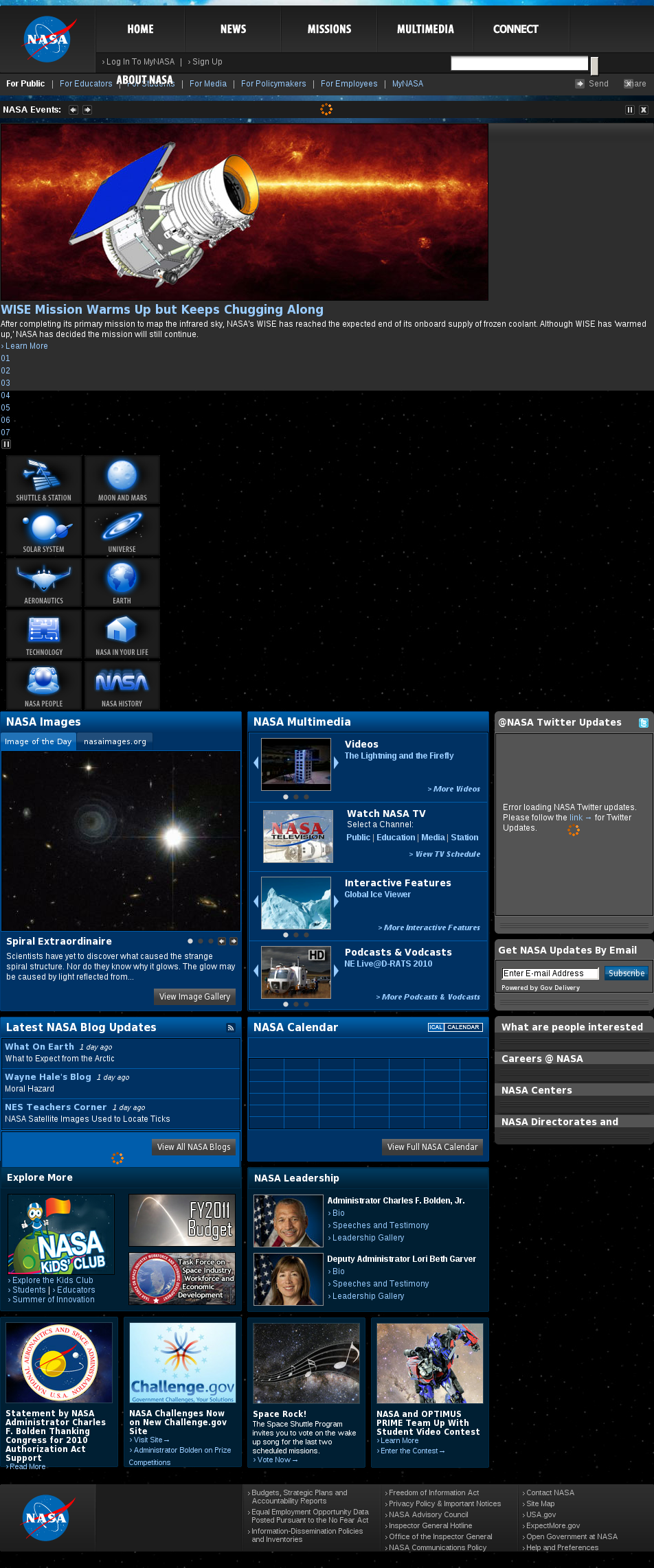}
}
\subfigure[2011]{
	\label{fig:nasa2011}
	\includegraphics[width=0.08\linewidth,natwidth=818,natheight=1915]{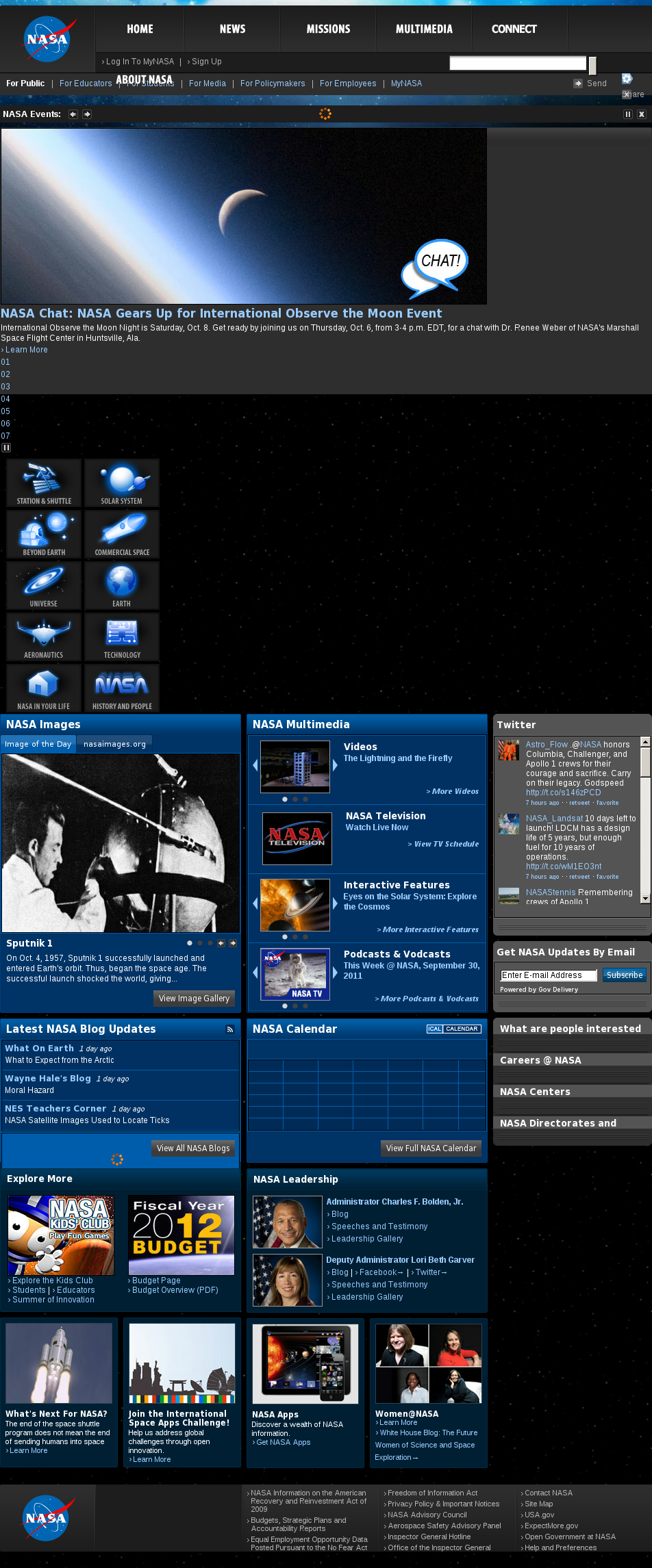}
}
\subfigure[2012]{
	\label{fig:nasa2012}
	\includegraphics[width=0.08\linewidth,natwidth=818,natheight=1915]{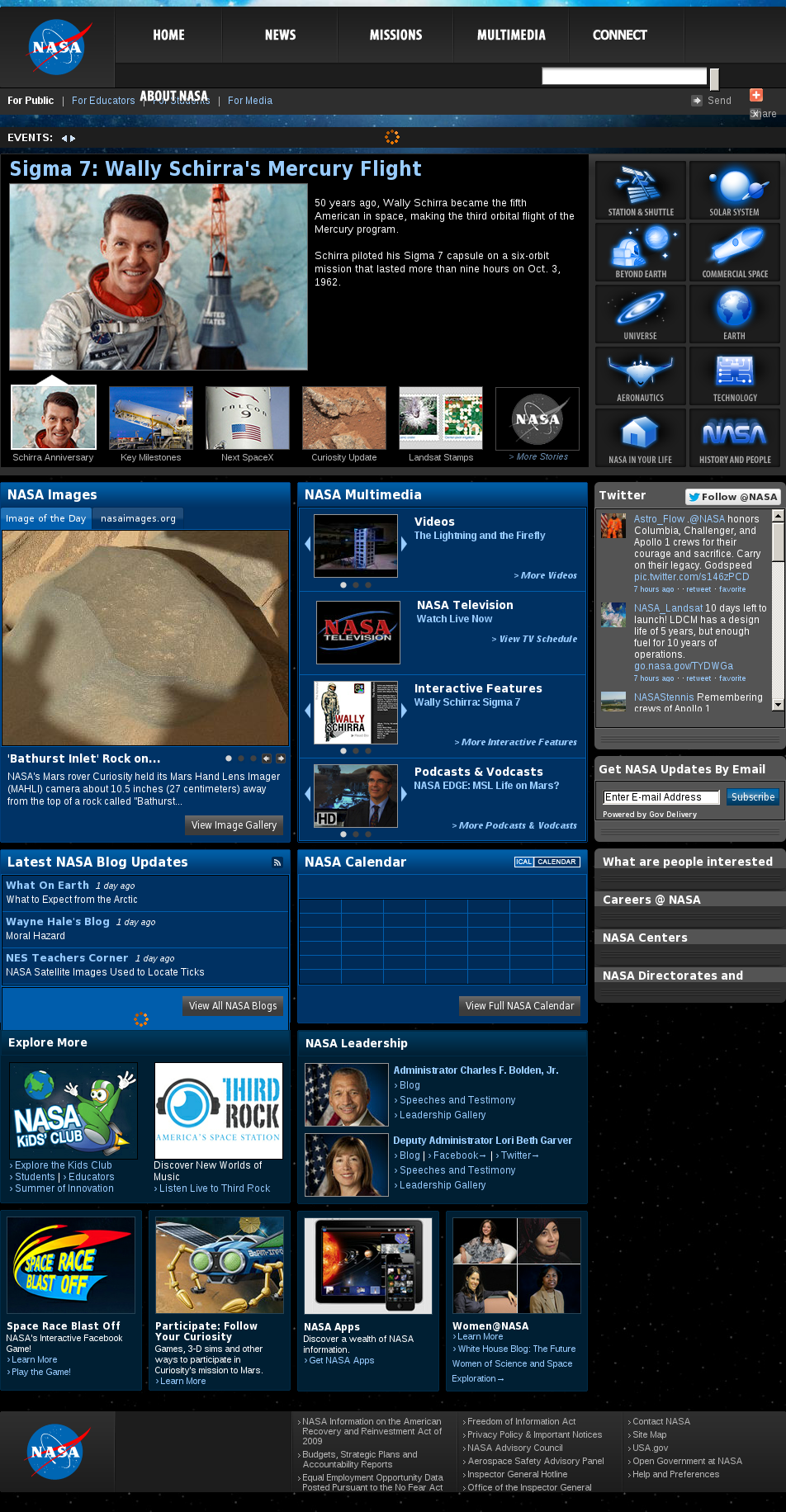}
}

\end{center}
\caption{NASA over time. Changes in design and thus the technologies used is easily observable between Figures~\ref{fig:nasa1997} and \ref{fig:nasa1998}, \ref{fig:nasa2002} and \ref{fig:nasa2003}, \ref{fig:nasa2006} and \ref{fig:nasa2006}, and \ref{fig:nasa2007} and \ref{fig:nasa2008}}
\label{fig:nasa}
\end{figure*}

A second example where the change in archivability over time is much more dramatic can be found in the NASA website\footnote{\url{http://www.nasa.gov}}. As a government funded agency, \\
 is advised to comply with the aforementioned accessibility standards. The same procedure (Section~\ref{sec:expSetup}) of creating a collection of annual mementos was used to obtain screenshots (Figure~\ref{fig:nasa}), HTTP logs, and the HTML of the memento. Mementos ranging from 1996-2006 were available and retained, a sampling that sufficiently spanned the introduction of dynamism into the web.

\begin{figure}[h]
\begin{center}
\subfigure[]{
\begin{lstlisting}
<script language="javascript" type="text/javascript" src="flash.js"></script>
<script language="javascript" type="text/javascript">
function flashURL(id) {
  //Flash Redirect URL
  window.location.href = 'index.html';
}
</script>
\end{lstlisting}
	\label{fig:nasa2003markup1}
}
\subfigure[]{
\begin{lstlisting}
var fstr = '';
if(hasFlash(6)) {						  
	fstr+='<object id="screenreader.swf">...</object>';
	window.status = 'Flash 6 Detected...';
} else {
	fstr+='...To view the enhanced version of NASA.gov, you must have Flash 6 installed....';
}
with(document) { open('text/html'); write(fstr); close(); }
\end{lstlisting}
	\label{fig:nasa2003markup2}
}
\caption{In 2003, \url{nasa.gov} introduced code (abbreviated here) into their website that checked the capability of the user's web browser and showed or hid content. The link to enter the website regardless of the user's browser capability, here, is generated with JavaScript. This would cause the content to not be displayed were the user's browser incapable or if client-side scripting were disabled in the user's browser preferences.}
\label{fig:nasa2003markup}
\end{center}
\end{figure}

\begin{figure}[h]
\begin{center}
\subfigure[]{
	\label{fig:plot}
	\includegraphics[scale=0.30,natwidth=610,natheight=642]{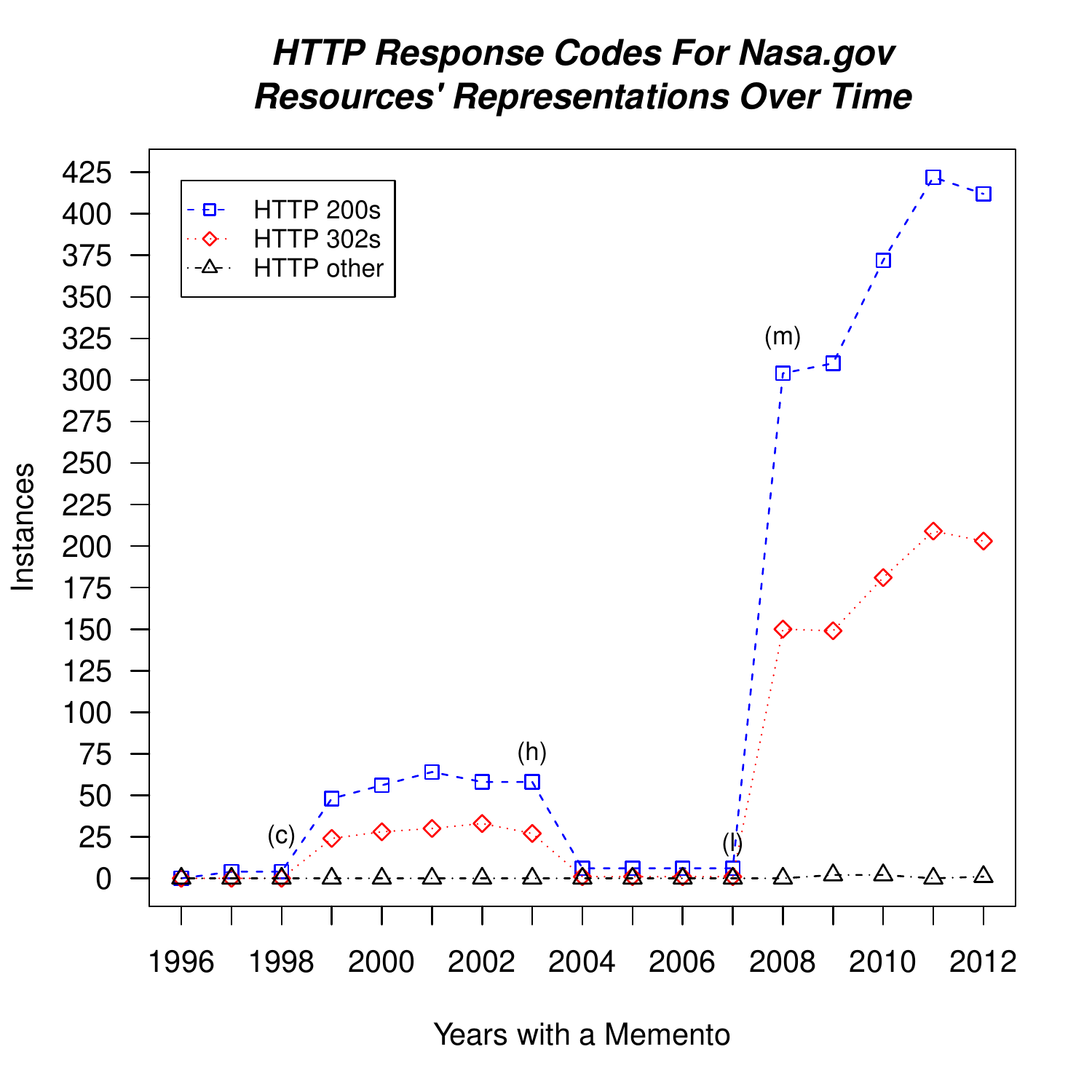}
}
\subfigure[]{
	\label{fig:plotwh}
	\includegraphics[scale=0.30,natwidth=610,natheight=642]{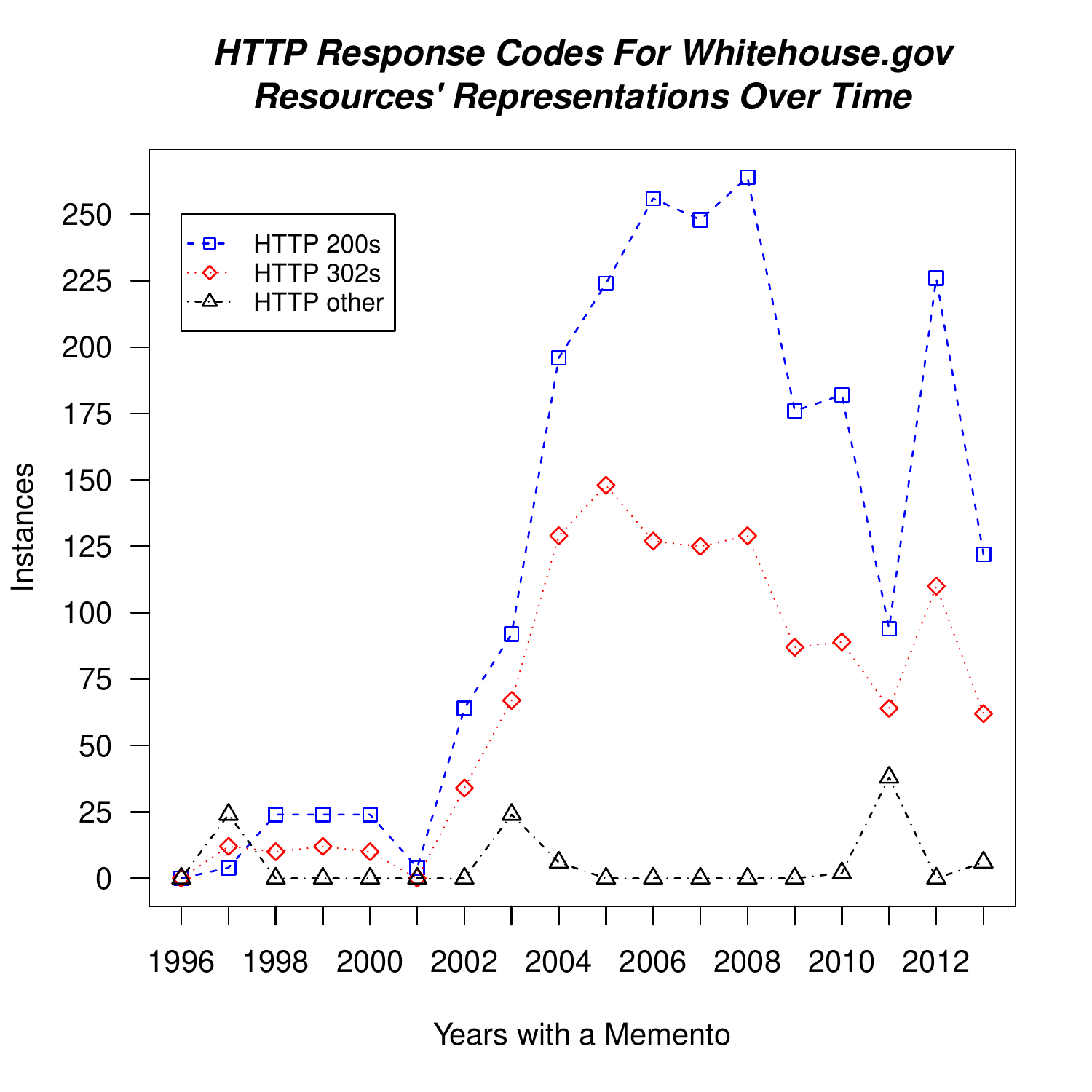}
}	
\caption{The number of resources required to construct the page (Figure~\ref{fig:plot}) has a noticeably absent lull that corresponds to Figure~\ref{fig:nasa}. The preservation of the White House web page (Figure~\ref{fig:plotwh}) exhibits a different problem yet is briefly similar in that the count drastically changed. The sudden change in 2011 is the result of a set of CSS files not reaching the crawler horizon, which may have had implications on subsequent resource representations (embedded within the CSS) from being preserved.}
\end{center}
\label{fig:graphs}
\end{figure}

The mementos from 1996 through 2002 show table-based websites devoid of JavaScript. In 2003, JavaScript was introduced into the markup (Figure~\ref{fig:nasa2003markup1}). Checkpoint 6.3\footnote{\url{http://www.w3.org/TR/WCAG10/wai-pageauth.html\#tech-scripts}} of the Web Content Accessibility Guidelines \cite{wcag} mandates that pages remain usable when programmatic objects are present on the page but not necessarily supported by the user. This is to ensure all content on a page is accessible.
Lack of accessibility directly correlates with unarchivability. Normally, providing an alternate means of viewing the page's content would suffice were the link to ``Enter NASA'' regardless of the incompatibility, but even the single relevant link on the page (with the other being a link to install Flash) is generated by a script (Figure~\ref{fig:nasa2003markup2}). From the 2004 to 2006 snapshots on, in lieu of testing for Flash, the ability to progress into the site is no longer offered but rather a message stating that JavaScript is required and a means to access instructions to enable it is the sole content supplied to the crawler. Observing the count of the resources required to construct a memento (Figure~\ref{fig:plot}) gives further evidence that both accessibility and archivability suffered between 2004 and 2006.

Relying solely on the number of resources fetched to determine where a site's reliance on unarchivable technologies lies is not foolproof, but it is a good guide to identify problematic pages. Were a crawler to encounter this drastic change and if the changed count was sustained, this should be noted as evidence of potential problems. On a comparable note, the same procedure was run on another government website where this deviation from web standards would be the least likely to surface, \url{whitehouse.gov}. A similar dip can be seen in Figure~\ref{fig:plotwh} in 2010. Examining the screen shot and log of HTTP codes, it is evident that a subset of CSS files were not preserved by the crawler. A preserved web page resembling this problem is not one necessarily related to the crawler's inability to fetch components of a page embedded in JavaScript but rather, the URI was not persistent enough to endure the time to reach Heritrix's horizon (the point at which it is preserved) once placed on the frontier (list of URIs to be preserved).

\section{Conclusions}

The archivability of websites has changed over time in different ways for different classes of websites. While JavaScript is partially to blame for this, it is more a problem that content is not accessible. Lack of accessibility makes content more difficult for crawlers to capture. Websites that are trend leaders, unfortunately, set a bad precedent for facilitating archivability. As this trend continues, tools are being created (Heritrix 3) that are archiving inaccessible websites. Recognizing techniques to make the archiving process easier by those that want their content preserved is a first step in guiding web development practices into producing web sites that are easier to preserve.

\bibliographystyle{abbrv}
\bibliography{sigproc} 

\end{document}